# The Synthescope: A Vision for Combining Synthesis with Atomic Fabrication

*Ondrej Dyck[1], Andrew R. Lupini[1], Stephen Jesse[1]*

[1] *Center for Nanophase Materials Sciences, Oak Ridge National Laboratory, Oak Ridge, TN*

**Abstract**

The scanning transmission electron microscope, a workhorse instrument in materials characterization, is being transformed into an atomic-scale material manipulation platform. With an eye on the trajectory of recent developments and the obstacles toward progress in this field, we provide a vision for a path toward an expanded set of capabilities and applications. We reconceptualize the microscope as an instrument for fabrication and synthesis with the capability to image and characterize atomic-scale structural formation as it occurs. Further development and refinement of this approach may have substantial impact on research in microelectronics, quantum information science, and catalysis where precise control over atomic scale structure and chemistry of a few "active sites" can have a dramatic impact on larger scale functionality and where developing a better understanding of atomic scale processes can help point the way to larger scale synthesis approaches.

**Introduction**

The scanning transmission electron microscope (STEM) stands poised for a revolution in function. Historical practice has cemented this instrument as a core analytical tool for the examination of materials down to the atomic scale. Changing the sample with the beam has historically been viewed negatively as beam damage and great pains have been taken to mitigate such effects. However, recent years have seen an interest in leveraging and controlling beam damage to bring about desired material transformations that have been demonstrated down to the atomic scale.[1–4]

Scanning probe microscopies (SPM) have been making strides toward atomically specified fabrication approaches for many years. These efforts began with the ground breaking demonstration of spelling out the letters "IBM" with single atoms[5] and have steadily advanced toward the application of this capability in the structuring of devices.[6,7] The STEM-based approach to material modification is at the same time both easier and more difficult. Sample transformations frequently occur unintentionally, illustrating the ease with which material modification can be accomplished. It is the controllability of the transformation that has remained a challenge and most efforts have been expended in the direction of *preventing* sample transformations. However, the successful mitigation of unintended effects has begun to bring a level of controllability to the STEM-based approach and the ease with which transformations can be produced (generally uncontrolled) lends credence to the idea that there is untapped potential for a useful fabrication technology.

These developments have begun to show that the STEM can be used for atomic scale manipulation. Indeed, the resolution of the instrument leaves no doubt that single atoms can be individually addressed, the challenge is determining what the beam can and cannot do in a seemingly infinite supply of different materials which may each have their own unique response to the influence of the beam. To this mixture we add numerous controllable parameters that are likely to have some bearing on the transformation process: sample temperature, beam current, accelerating voltage, and sample environment.

Some of the authors of this work have previously laid out a concept dubbed the "Atom Forge".[8]

In this conceptualization the STEM functions as a platform capable of material manipulation at the atomic scale, incorporating various feedback control signals to direct the workflow through real-time data analytics and decision making. While such a vision is certainly ambitious it is also necessarily vague on many points, notably, which materials, and which transformations. Given the range of possibilities it is not obvious which directions to pursue.

Here, we lay out a next step in the natural evolution of the field, namely the incorporation of synthesis processes in the STEM. The objective here is not just to perform a particular synthesis task but to facilitate the discovery of new potential transformations. Most experiments that leverage STEM-based atomic scale material transformations rely on crystallization,[9,10] amorphization,[10] atom conserving atomic movement,[10–17] or atomic ejection (milling/sculpting).[18–26] (Notably, these transformations often involve a substantial transfer of energy to the sample that is not available in SPM-based approaches but they typically lack the single atom controllability.) Fewer studies examine the addition of material atom-by-atom *in situ*. This bias is partially due to the difficulties involved in supplying material to the sample – termed the 'workpiece' in the context of this discussion – in the vacuum system of the microscope but perhaps also lack of an atomic level conceptualization of the outcome one desires. For example, within the 3D printing paradigm materials are "deposited" on a substrate and the adherence of new layers is brought about typically by cooling the source material to induce a phase change and restructuring to facilitate strong chemical bonding. The types of chemical bonds are clearly important, but the specific nature of each bond is almost irrelevant. What matters, rather, is the ensemble behavior. However, as such a process is scaled down, the importance of each bond in the behavior of the ensemble increases until, at the level of the single atom, there is no ensemble, only the bonding of one atom to the substrate. This is where we must be clearer about what "deposition" is intended to mean for the atoms involved.

Materials synthesis is a broad category generally defined as the combining of precursor materials to form a new substance. The strategies for accomplishing this are similarly broad (e.g. spontaneous chemical reactions, thermally induced reactions, mechanical mixing, optical excitation, etc.), however, the synthesis process occurs only within the confined region of space defined by the presence of the necessary ingredients. Synthesis is inherently a local phenomenon. Here, we would like to consider this property of locality and attempt to lay out a vision for scaling synthesis to the level of a single atom.

**A Synthesis Chamber the Size of an Atom**

Within the STEM we have a controlled environment (vacuum) with a focused beam of electrons which creates a highly localized perturbation of the sample. We can consider this localized perturbation as our synthesis "chamber"—or, more precisely, it is a chamberless synthesis environment (CSE) the extent of which is defined by the intersection of the beam and the substrate. Within this perturbed region of space, the sample environment is drastically different from regions outside the perturbation. There are electronic excitations, the emission of electrons, momentum transfer, ejection of atoms, changes to chemical potentials, etc, which collectively determine the evolution of the structures within that region. A serious drawback is that these influences cannot be separated; for example, the e-beam excitation cannot be controlled to the extent that it excites a sample electron but not also simultaneously transfer momentum to the nucleus. Moreover, one usually cannot pause some transformation at a metastable point in the transformation process and change a critical parameter to direct the outcome. Once the system is set in motion, it will evolve accordingly to the next stable state. Atomic scale transformational process typically occur on a much shorter time scale than one can detect or respond to. These constraints notwithstanding, there still exist many options for exercising control over the structural and chemical evolution within the CSE.

What the STEM is specifically designed to do is to focus the beam of electrons (in many cases to less than an angstrom) and also *move* it across the sample. For the purposes of the discussion here, this is a pivotal point. The CSE can be moved around with great spatial precision and encounter a variety of different physical systems in a non-uniform sample. Likewise, from the perspective of a fixed position on the sample, the CSE can be turned on and off by unblanking and blanking the beam, or by positioning it elsewhere on the sample. In this way, a reaction process could be prepared by adjusting global parameters (e.g. partial pressures, temperature, electric or magnetic field strength and orientation, optical excitation) without the intended process occurring until the CSE is created through e-beam exposure.

**From the Molecule to the Atom**

*Molecule-by-Molecule Deposition*

We can view e-beam induced deposition (EBID) from the CSE perspective. Global parameters—the substrate, the precursor partial pressure, the temperature, the e-beam fluence and accelerating voltage—are set before the deposition begins and once these parameters have been established, the CSE is manifested by e-beam irradiation and moved to whatever location the user wants to induce the chemical reaction. This process is typically described as the dissociation of the precursor gas by the e-beam and the subsequent attachment of the chemically reactive molecular fragments to the substrate. van Dorp et al performed a series of pioneering investigations aimed at transferring the EBID concept from the scanning electron microscope (SEM) to STEM[27–32] and, as a consequence, reap the enhancement in resolution afforded by the finer STEM e-beam profiles and decreased substrate scattering in thin samples (which limits resolution in SEM). Ultimately, they were able to achieve molecule-by-molecule deposition. Their landmark results, published in 2012, are summarized in Figure 1.[30]

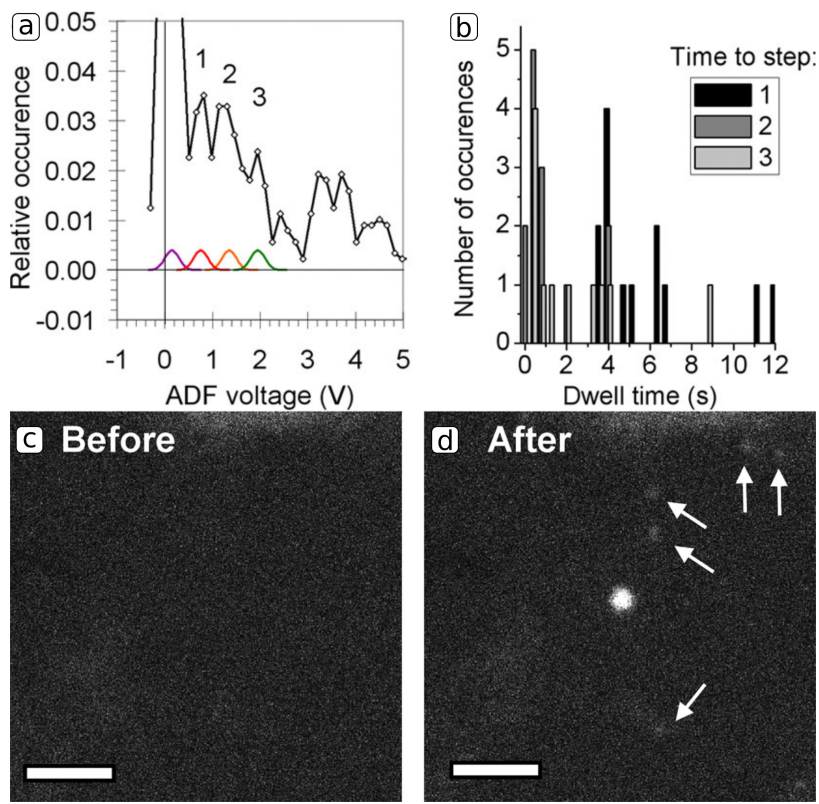

**Figure 1 Summary of molecule-by-molecule deposition reported in 2012 by van Dorp et. al.** (a) analysis of the observed voltage reading on the ADF detector showing discrete intensity levels indicating the quantized addition of material. (b) Histogram of the distribution of occurrences of the first three peaks labeled in (a). (c) Before and (d) after images of the deposition site. Arrows mark unintended deposition from precursor fragments. Scale bars represent 5 nm. Reproduced with permission from reference [30].

Careful examination of the ADF voltage output reading, Figure 1(a), reveals a stepwise increase indicating discrete additions of material to the deposition site. From this information, the authors were able to infer that this represented integer multiples of molecular fragments attaching to the sample surface and, thus, demonstrated molecule-by-molecule deposition. The histogram shown in Figure 1(b) displays the length of time required for the peaks 1-3 to appear. Deposition of the first molecule required on average 5.6 s while the attachment of the second and third molecules required 0.9 and 2.1 s respectively. The authors attribute this to a longer residence time of the precursor on an already formed deposit as compared to the pristine graphene surface. The before and after images are shown in Figure 1(c) and (d), respectively. From the resolution shown here, we can see that the authors are likely not able to inquire further about the deposition process, e.g. the atomic structure of the deposition site as each molecule is attached. From the EBID perspective as well, the precise bonding configurations are usually ignored as they are, to a large degree, irrelevant.

It is well known that EBID deposits are often contaminated with carbon remnants from the dissociated precursor material and hence the authors describe this process as "molecule-by-molecule" rather than "atom-by-atom". At this scale, the atomic structure and chemical bonding of each atom begin to become significant. These results make clear that for progress to be made on this front a deeper understanding of the atomic processes involved in deposition are required.

*Atom-by-Atom Deposition*

A naive approach toward addressing the carbon remnants would be to simply supply pure source material. van Dorp et. al. used $W(CO)_6$ as their precursor which contains six times more carbon atoms than tungsten atoms. However, the CO is essential to make this molecule a volatile gas species. Without them we would have the notoriously thermally robust tungsten metal (with a boiling point around 6000 K). Moreover, if we were to vaporize and deliver isolated tungsten atoms to our region of interest, there would be no molecule for the e-beam to dissociate and therefore no localized chemical bonding defined by the e-beam. There is, nonetheless, a way forward; use the e-beam to create attachment points in the substrate material while the purified atoms are migrating along the surface.

This strategy was developed by the authors of this article and presented in several publications as its viability for the attachment of dopant atoms in graphene was assessed.[16,17,33–35] Later developments more clearly drew the connection to the scaling of EBID to the level of the atom.[36,37] In Figure 2, we highlight one result where Cu atoms were patterned in twisted bilayer graphene using the e-beam to create attachment points. The temperature of the sample was adjusted to control the availability and mobility of the Cu atoms, which were sourced from Cu nanoparticles pre-deposited on the sample surface. When the sample temperature was elevated to approximately the melting temperature for bulk Cu, the CSE transitions from producing holes (caused by the momentum transfer from the beam to carbon atoms) to producing holes that are rapidly occupied by Cu dopants and dopant clusters, shown in Figure 2(d). Therefore, through a combination of finely tuned global and local conditions and processes, we can move from creating arrays of atomic scale holes[38] to creating arrays of precisely positioned dopants.

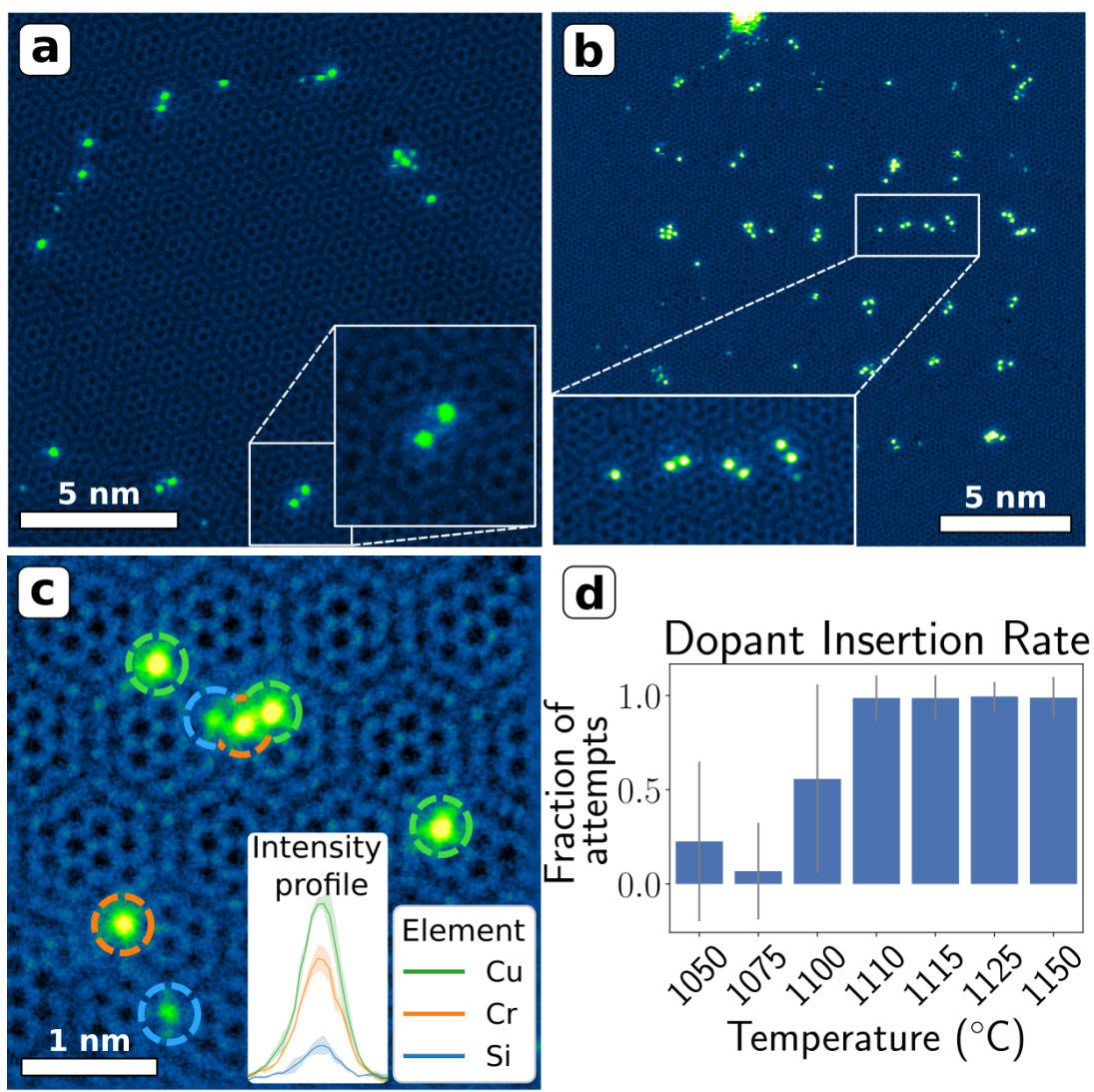

**Figure 2 Summary of Cu dopant patterning in TBG.** (a) Example circle and (b) array patterned with Cu dopant atoms. (c) Magnified view of an area where three different elemental species were incorporated adjacent to one another. (d) Success rate of getting a dopant (vs a hole) as a function of substrate temperature. Adapted with permission from reference [36].

**Assessing the Trajectory of the Field**

Here, we summarize more explicitly the strategic progression. The EBID approach to direct deposition is well established and is mostly applied using SEM and FIB instrumentation. In scaling this approach toward the level of single atoms using a STEM, van Dorp et al[30] highlight some of the challenges with this approach. In particular, the level of cleanliness required to achieve the molecule-by-molecule deposition shown in Figure 1 is much more stringent than typically employed for EBID in an SEM. In the supplementary information of reference[30] van Dorp et al detail their efforts at reducing extraneous hydrocarbon availability including prolonged (24 h) plasma cleaning, operating at elevated temperatures (800 °C), and employing the use of a cryogenic anti-contamination device. Ultimately, they found that working with small, few-layer graphene flakes was necessary to mitigate hydrocarbon diffusion into their region of interest (large-area single-layer graphene promotes long-range hydrocarbon surface diffusion). Nevertheless, their precursor molecules, $W(CO)_6$, themselves have significant carbon content. Koops et al.[39] estimate the deposited material typically contains 55 at. % tungsten, 30% carbon, and 15% oxygen. Figure 3(a) illustrates deposition using the EBID approach, assuming a perfectly clean deposition environment with the deposited material comprised of a mixture of the elements from the source gas. The nature of this approach to deposition is inherently limited by the dissociated molecular fragments. In other words, this approach cannot be improved from molecule-by-molecule deposition to atom-by-atom deposition without some fundamental change being made to the process itself.

To achieve atom-by-atom deposition, the precursor material must be dissociated into single atoms. However, once the precursor has been atomized, there are no further chemical alterations that the e-beam can induce. The mechanism for attachment to the substrate in the EBID approach is the e-beam induced chemical and bond change to the precursor molecule, followed by the chemical bonding of the molecular fragments to the substrate close to the e-beam position. Instead of altering the precursor atoms, the e-beam can be used to selectively alter the substrate at desired locations. This technique was originally developed to facilitate the insertion of Si dopant atoms into graphene for further atomic manipulation experiments.[16,17] It was later generalized and expanded to include many other elements[33–35] and is likely the mechanism underlying earlier experiments showing the separation of small atomic clusters on graphene using the e-beam.[40] This approach, summarized in Figure 3(b), uses the e-beam to create vacancies/defects in the substrate material through e-beam damage. These vacancies/defects are chemically reactive relative to the surrounding pristine material as a result of the undercoordinated atoms. The e-beam is then moved onto the source material (typically a nanoparticle) to sputter individual atoms onto the substrate surface. Sputtered atoms that diffuse to the vacancy locations will, with high likelihood, chemically bond to the defect site. The substrate can undergo restructuring processes as well, depicted in Figure 3(b)(iii), which may mitigate the reactivity of the attachment points. The transformation pictured is the conversion of two single vacancies, collectively harboring six undercoordinated carbon atoms, into a single reconstructed divacancy structure with no undercoordinated carbon atoms.

This strategy typically requires repeated iterations of the procedure (i)-(iii) before successful attachment of an atom is accomplished. This approach, therefore, does not offer a continuous, direct-write process like the EBID strategy shown in Figure 3(a). However, it highlights the atomic level of detail needed for an understanding of how to move from the molecule to the atom. For example, even the depiction of e-beam deposited material shown in (a) is simply a cluster of randomly positioned spheres, while the captured atom in (b)(iv) is judiciously positioned at an exact location, *reflecting the much more specific understanding of how bonding is accomplished in this case*.

One of the drawbacks to the approach depicted in Figure 3(b) is the requirement that the e-beam perform both vacancy generation and sputtering/atomization of the source material. These two processes are not performed at the same time because the single e-beam is moved between the two locations. Moreover, to supply numerous sputtered atoms one would like to have high beam currents and primary energies, but to maintain control over the defect generation process one would like more moderate beam currents and energies. In addition, there can be a significant time delay between the defect generation and the arrival of a source atom, giving the substrate ample time to restructure. If atomized material can be supplied without the need for e-beam sputtering these processes could be performed at the same time, representing a significant improvement. Such a strategy was presented in Figure 2 and is schematically represented in Figure 3(c).[36] In this approach, the sample is heated until the source material, in the form of nanoparticles sitting on the sample surface, begins to atomize and diffuse across the substrate. The substrate temperature also plays a role in promoting surface diffusion. Using this approach, arbitrary patterns of single and few atom clusters could be attached to the twisted bilayer graphene substrate.

The dopant insertion success rate, shown in Figure 2(c), illustrates a pronounced role for the temperature. However, because the source material is physically supported by the substrate, it is not possible to independently investigate the role played by source evaporation rate vs. surface diffusion. To properly investigate these processes, the source material should be removed from the substrate as depicted in Figure 3(d). This has the added advantage of not requiring extraneous material physically attached to the surface which will likely be undesirable in most fabrication applications.

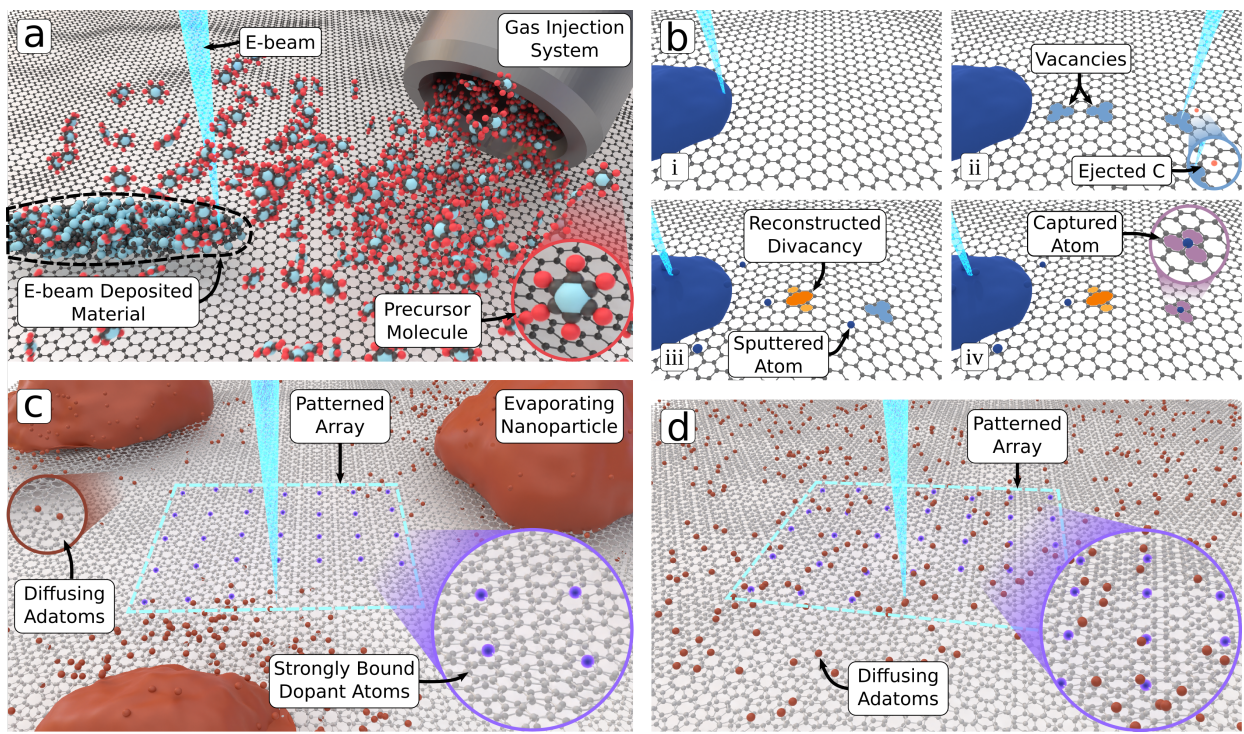

**Figure 3 Strategic progression of the field toward atomic fabrication.** (a) The EBID approach to e-beam deposition performed by van Dorp et al.[30] (b) The cyclic damage and sputtering approach developed by Dyck et al.[35] where the steps i-iv are repeated in succession until dopant insertion is accomplished. (c) Long range dopant insertion facilitated by evaporating nanoparticles on the graphene surface, performed at elevated temperatures.[36] (d) Removal of the source material from the sample surface will enable independent control over the source evaporation rate and the substrate temperature. Additionally, this approach preserves the cleanliness of the sample surface.

The picture depicted in Figure 3(d) begins to look like a special case of synthesis where the chemical reactions are promulgated by beam-sample interactions at specifically chosen positions. The combination of these approaches (synthesis and STEM) represents a radically different conception of both how synthesis is performed and what is possible in the STEM. Here, we envision synthesis capabilities that are tuned to initiate a reaction at predefined e-beam locations. Further evolution of the system could proceed via the tuning of global parameters (e.g. substrate temperature, precursor availability) and snapshots of the process viewed directly as the growth occurs. Moreover, this strategy lends itself to real-time feedback control where information about the current sample state can be used to dynamically alter the growth parameter landscape. This could include timed introduction of new nucleation sites, switching of elemental source materials, adjustments to various temperatures (substrate or source), optical excitation, electrical biasing, and field gradients. Combined with the already powerful set of analytical characterization modalities, such an instrument would break new ground in unveiling dynamic atomic processes. At the same time, combining synthesis strategies may provide a methodology to address the inherent slowness problem of top-down sequential processing. Bottom-up growth processes would be leveraged whenever feasible, reserving the tedious top-down fabrication only for certain, site-specific tasks.

**Realistic and Practical Pathways Forward**

Unfortunately, STEM instruments are not currently arrayed with many commercial options for *in situ* synthesis. Nevertheless, here we present several example concept drawings that leverage and improve upon existing equipment with minor modifications. The primary purpose here is to furnish concrete illustrations of how the combination of several existing capabilities with some relatively simple modifications could prime the field for demonstrations of this kind. The concept drawings are based on the Nion cartridge design, with which the authors are most familiar, however it is assumed that these designs would be similarly adaptable to other microscope platforms.

The first concept drawing, shown in Figure 4(a), illustrates the incorporation of a resistive heating element onto an existing electrical cartridge originally designed to facilitate *in situ* heating, biasing, and transport experiments. In this drawing, the removable insert of the original design has been replaced with a base plate and printed circuit board (PCB) which holds the device/sample at the center and is outfitted with a heater element on the end. The heater element is envisioned to be coated with the source material that can be evaporatively delivered to the device/sample through externally applied voltages. At the same time, the device/sample can be electrically connected through the PCB-mounted device contact block. This will enable the operation of on chip local heating for independent control of surface diffusion, application of local electric fields, and/or real-time monitoring of transport properties.

An alternative approach to material delivery is to use *in situ* laser ablation, illustrated in Figure 4(b). In this concept, a laser would be used to ablate nearby targets affording the option to rapidly change source materials. For our cartridge design this would require the removal of a portion of the cartridge rail to provide line-of-sight access from the laser to the target, as well as an ablation target support block on the opposite side of the cartridge.

A series of magnified images of the device/sample are shown in Figure 4(c)-(e) which harbors an electron transparent membrane at the center to facilitate high resolution STEM characterization and manipulation. The particular design shown is based on the authors' prior work,[41] however custom device fabrication for experimentation is increasingly common[42–45] and STEM compatible integrated circuit designs have also been pursued with success by other groups[20–22,46–49] suggesting the general applicability and attractiveness for operando investigations.

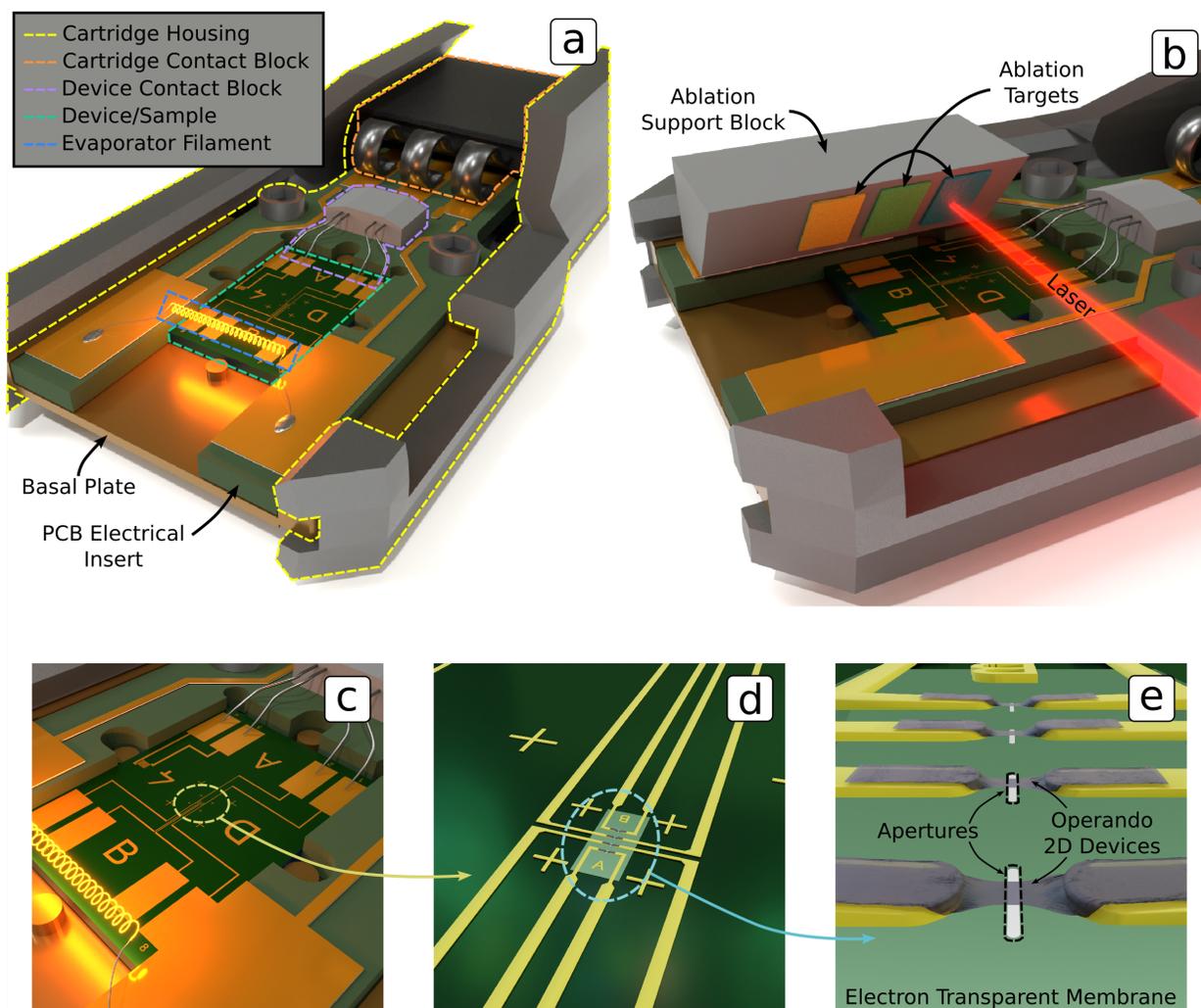

**Figure 4 Concept drawings illustrating simple approaches toward the development of the synthescope approach.** (a) An evaporator filament is incorporated on the end of a printed circuit board (PCB) that overhangs the device/sample. Electrical operation is facilitated through the use of existing electrical contacting holders. (b) A strategy for laser ablation deposition is illustrated. This approach would require cutting an access notch in the holder cartridge and installation of a commercially available laser. Multiple ablation targets could be used to facilitate switching source materials. (c) Magnified view of (a) showing electrical integration with a previously developed wafer-scale chip design. (d) Electrical traces converge at the center of the chip where the operational devices are supported on an electron transparent membrane (e) to be compatible with STEM imaging.

## Speculative Future Directions

With a platform of the kind suggested, synthesis processes could be performed simultaneously with device operation, direct, atomically resolved characterization, and alteration down to the level of single atoms. It is well known that e-beam damage can remove material enabling precise sculpting and it has been shown in some cases that these kinds of modifications can be performed with atomic specificity.[18,24–26,50–54] These examples represent a highly controlled removal of material and several studies have coupled this capability with operando devices to explore the effects of architectural modification in real time.[20–23,47]

With the addition of synthesis capabilities, we envision the ability to begin *adding* material at precisely defined locations. By damaging the substrate, shown in Figure 2, we can see that spontaneous chemical reactions fix impurity atoms to the lattice. These might be the intended final structure or they might form the nucleation sites for further processing. For example, it might be desirable to position a specific impurity element at the nodes of a TBG membrane to imbue it with new functionality. This goal appears to be within reach of current capabilities.

Alternatively, one might supply additional atoms to promote the growth of larger 2D or 3D structures from these initial impurity positions. Discovery of the correct processing conditions may lead to the ability to precisely position, for example, plasmonic arrays of nanoparicles or quantum dots. If layer-by-layer growth can be achieved and controlled, it is also conceivable that a particular defect or impurity could be created at a specific position in each layer resulting in atomically specified alterations being accomplished in the third dimension—3D printing of defects during the growth process.

Moreover, changing the source material, as is implied by the multiplicity of ablation targets in Figure 4(b), would allow one to alternate between different growth chemistries on the fly while directly monitoring the outcome. One could imagine growing 2D and 3D crystals while cycling the element being deposited and having feedback-controlled timing of the process based on direct observation of the structure being formed. In addition, patterning alternating impurity elements at the nodes and antinodes of a twisted bilayer 2D material should be possible.

It possible that the influence of the e-beam itself could be leveraged to drive and guide the synthesis processes. Indeed, this is the underlying idea behind several demonstrations of e-beam guided transformations.[9,10,55,56] Having better control over the supply of the source material and a better understanding of the interaction with the e-beam could lead to atom-by-atom guiding of the growth process similar to the recent demonstration of atom-by-atom direct writing in the STEM.[37] This can be thought of as a directed growth process driven by real-time modification of the chemical potential of the substrate using the e-beam as a CSE. The range of possibilities and vast parametric landscape of unexplored combinations is both daunting and exciting. In every direction lie potential new discoveries and physical and chemical phenomena to be refined into fabrication workflows.

**Conclusion**

The concept drawings in Figure 4 are intended to provide concrete representations of possible ways to integrate synthesis capabilities with current microscope hardware, as well as to illustrate that such modifications could be relatively simple to implement as many of them are incorporated into the sample making it easy to insert or remove from the microscope and obviating the need for changes to the microscope itself. Of course, one is not limited to these strategies, but the overarching trend of operando devices, STEM-based manipulation, and increasingly clean and precise deposition of atoms suggests that some manifestation of the synthescope is on the scientific horizon. These developments have the potential to transform our understanding of material growth processes and atomic scale chemical interactions. Beyond observation, we will be able to influence synthesis *in situ* at the level of the single atom.

**Acknowledgement**

This work was supported by the U.S. Department of Energy, Office of Science, Basic Energy Sciences, Materials Sciences and Engineering Division (O.D. A.R.L., S.J.), and was performed at the Center for Nanophase Materials Sciences (CNMS), a U.S. Department of Energy, Office of Science User Facility.